\documentclass[12pt]{article}
\usepackage{graphicx}
\usepackage{color}
\usepackage{hyperref}
\usepackage{subfigure}
\usepackage{multicol}
\usepackage{epsfig}
\usepackage{caption}
\usepackage{wrapfig}

\newcommand\pubnumber{}
\newcommand\pubdate{\today}

\def\graz{Institute of Physics, University of Graz, \\
8010 Graz, Austria}
\def\tifr{Department of Theoretical Physics, \\
Tata Institute of Fundamental Research, Mumbai, India.}

\def\Title#1{\begin{center} {\Large #1 } \end{center}}
\def\Author#1{\begin{center}{ \sc #1} \end{center}}
\def\Address#1{\begin{center}{ \it #1} \end{center}}

\newcommand\pubblock{\rightline{\begin{tabular}{l} \pubnumber\\
         \pubdate  \end{tabular}}}
\newenvironment{Abstract}{\begin{quotation}  }{\end{quotation}}
\newenvironment{Presented}{\begin{quotation} \begin{center} 
             PRESENTED AT\end{center}\bigskip 
      \begin{center}\begin{large}}{\end{large}\end{center} \end{quotation}}

\newcommand\bef{\begin{figure}}
\newcommand\eef[1]{\label{fg:#1}\end{figure}}
\newcommand\bec{\begin{center}}
\newcommand\eec{\end{center}}
\newcommand\besf{\begin{subfigure}}
\newcommand\eesf[1]{\label{sfg:#1}\end{subfigure}}
\newcommand\beq{\begin{equation}}
\newcommand\eeq[1]{\label{#1}\end{equation}}
\newcommand\beqa{\begin{eqnarray}}
\newcommand\eeqa[1]{\label{#1}\end{eqnarray}}
\newcommand\bet{\begin{table}}
\newcommand\eet[1]{\label{tb:#1}\end{table}}
\newcommand\best{\begin{subtable}}
\newcommand\eest[1]{\label{stb:#1}\end{subtable}}
\newcommand\betb{\begin{center}\begin{tabular}}
\newcommand\eetb{\end{tabular}\end{center}}
\newcommand\beit{\begin{itemize}}
\newcommand\eeit{\end{itemize}}

\newcommand{\cref}[1]{\bec  \textcolor{gray}{\scriptsize #1}\eec}

\newcommand\fgn[1]{Figure \ref{fg:#1}}
\newcommand\eqn[1]{eq.\ (\ref{#1})}

\newcommand\incfig[2]{\includegraphics[scale=#1]{./#2}}

\definecolor{DarkGreen}{rgb}{0.00,0.29,0.00}
\definecolor{DarkRed}{rgb}{0.79,0.00,0.00}

\def\prsp#1#2%
  {\mathop{}%
   \mathopen{\vphantom{#2}}^{#1}%
   \kern-\scriptspace%
   #2}

\begin{document}
\begin{titlepage}
\pubblock

\vfill
\Title{Charmed baryons on the lattice}
\vfill
\Author{M. Padmanath\footnote{Speaker}}
\Address{\graz}
\vfill
\Author{Nilmani Mathur}
\Address{\tifr}
\vfill
\begin{Abstract}

We discuss the significance of charm baryon spectroscopy in hadron physics and review the recent 
developments of the spectra of charmed baryons in lattice calculations. Special emphasis is given 
on the recent studies of highly excited charm baryon states. Recent precision lattice measurements 
of the low lying charm and bottom baryons are also reviewed.

\end{Abstract}
\vfill
\begin{Presented}
The 7th International Workshop on Charm Physics (CHARM 2015)\\
Detroit, MI, 18-22 May, 2015
\end{Presented}
\vfill
\end{titlepage}
\def\thefootnote{\fnsymbol{footnote}}
\setcounter{footnote}{0}
%

\vspace{-0.3cm}
\section{Introduction}
Charmed hadrons play an important role in understanding the dynamics of QCD, the theory of 
strong interactions. The presence of quarks with masses significantly greater than $\Lambda_{QCD}$ 
provides a flavor tag, which could help in understanding the mechanism of confinement and 
the systematics of hadron resonances that are obscure due to the chiral dynamics in 
light baryons. Charmonia and charmed mesons have been studied comprehensively both 
theoretically and experimentally \cite{Daniel}. However, charmed baryons 
have received substantially less attention, although they can provide similar insight 
into the nature of QCD. 

{\bf The triply charmed baryon} is expected to be the ideal candidate for understanding the dynamics 
inside the baryons, as pointed out by Bjorken several years ago \cite{Bjorken:1985ei}. With no 
valence light quarks, the spectra of these baryons can reveal information on the interplay between 
the non-perturbative and perturbative effects, and thus can elucidate our knowledge about the nature 
of strong interactions. On the theoretical side, the constituent quark models are expected to 
describe the properties of the low lying spectrum of triply charmed baryons. However, for the 
excited states this description becomes more unclear. There are no experimental observation of 
triply charmed baryons, although QCD predicts existence of such states.

The excited spectra of {\bf doubly charmed baryons} and their splittings can shed light into their 
intrinsic collective degrees of freedom, which are characterized by two widely separated scales : 
the low momentum scale of the light quark ($\sim \Lambda_{QCD}$) and the relatively heavy 
charm quark mass, giving rise to excitations in these systems. Only SELEX collaboration has 
reported discovery of doubly charmed baryons \cite{Mattson:2002vu}. However, these states have not 
been observed either by BaBar \cite{Aubert:2006qw}, Belle \cite{Kato:2013ynr} in $e^+e^-$ 
annihilation experiments, FOCUS in photo-production experiments \cite{Ratti:2003ez}, or very 
recently by LHCb at baryon-baryon collider experiments at CERN \cite{Aaij:2013voa}. The helicity angular 
distribution analysis by SELEX suggests that an isospin splitting of $\sim$17 MeV for the ground 
state $\Xi_{cc}^+$ baryon, which is unusually large in comparison with the numbers for light, 
strange and singly charmed baryons. Such a large isospin splitting could be explained by a larger 
Coulombic interaction than the strong interaction \cite{Brodsky:2011zs}. However, such interaction 
limits the size of doubly charmed systems to unusually compact dimensions. Thus there is no precise 
understanding of this unusually large isospin splittings observed for doubly charmed baryons.

{\bf Single charmed baryons} can guide the investigations of diquark correlations, 
the effective degrees of freedom to describe the dynamics inside baryons. In the light 
baryons, such investigations are difficult as the three diquark correlations in them follow
similar dynamics. In the presence of a charm quark, one expects the difference between
the diquark correlations within the light quark pair and heavy-light pair to reflect in 
the excited state spectra, decays, branching ratios and the production rates of charmed 
baryons \cite{Shirotori:2014nua}. Thus excited spectra of these systems can shed light into 
the collective modes in these systems and give insight on freezing degrees of freedom, if 
it happens at all, and the missing baryon resonances. Thus, a systematic study of these 
excitations as a function of quark masses 
can aid in understanding the energy spectra of baryons from light to bottom, including 
hyperons and charmed baryons. Out of $\sim$17 
known singly charmed baryons (with *** or more), the quantum numbers for only few excitations 
are known from experiments, while most of the assignments are quark model expectations 
\cite{Agashe:2014kda}. Prospects for singly charmed baryons also come from finite temperature 
lattice studies of partial pressures of heavy-light hadrons across the deconfinement 
crossover \cite{Bazavov:2014yba}. They find significant contribution from additional charmed 
baryon states over an estimate from experimentally known excitations.

The wealth of prospects from charmed baryon spectroscopy described above and the anticipated 
large sets of statistical samples being collected at dedicated experiments, at J-PARC, LHCb 
experiment, future PANDA experiment at the FAIR facility, Belle II at KEK and BES III, which 
could provide lots of information on charmed baryons, calls for a quantitative understanding of 
the charmed baryon spectra using non-perturbative first principles calculations such as 
lattice QCD. All results from such lattice calculations can provide crucial inputs to the 
future experimental discovery. Charmed baryons have also been studied theoretically by 
non-relativistic and relativistic potential models, effective field theories with potential 
NRQCD, heavy quark spin symmetry, etc. A review of such calculations can be found 
in the Ref. \cite{Crede:2013sze}. Lattice calculations would also be important to compare with the 
results from these effective field theories and understand the dominant interactions that 
gives rise to the observed spectra.

\vspace{-0.3cm}
\section{Low lying spectra}

Lattice QCD computations provide a powerful tool to perform ab-initio calculations of the QCD 
spectra and so to learn about the non-perturbative low energy regime of QCD. Over the past 
few years there has been remarkable progress towards simulations with physical quark masses, 
with many ground state hadron observables being measured with an impressive statistical precision 
and full control over the systematic uncertainties \cite{Aoki:2013ldr}. Lattice computation 
of hadron masses proceeds through the calculation of the Euclidean two point correlation 
functions between creation operators ($\bar{O}_i$) at time $t_i$ and annihilation operators ($O_j$)
at time $t_f$. 
\beq C_{ji}(t_f-t_i) = \langle
0|O_j(t_f)\bar{O}_i(t_i)|0\rangle = \sum_{n} {Z^{n*}_iZ^n_j\over 2  E_n} e^{-E_n(t_f-t_i)}, \eeq{2pt}
where $E_n$ is the energy of the $n^{th}$ excited state. The spectral information are extracted 
from the overlap factors ($Z^n_j$) for one or more two point correlation functions. Typical charmed 
baryon operators consist of three quark fields ($\epsilon_{abc}q^aq^bq^c$), with at least one of 
them being charm. Although all the lattice calculations follow this general procedure, there are 
important differences in the formulations that could lead to different systematic errors. Consistency 
between various formulations gives confidence in the results. 
\bef
\centering
  \incfig{0.8}{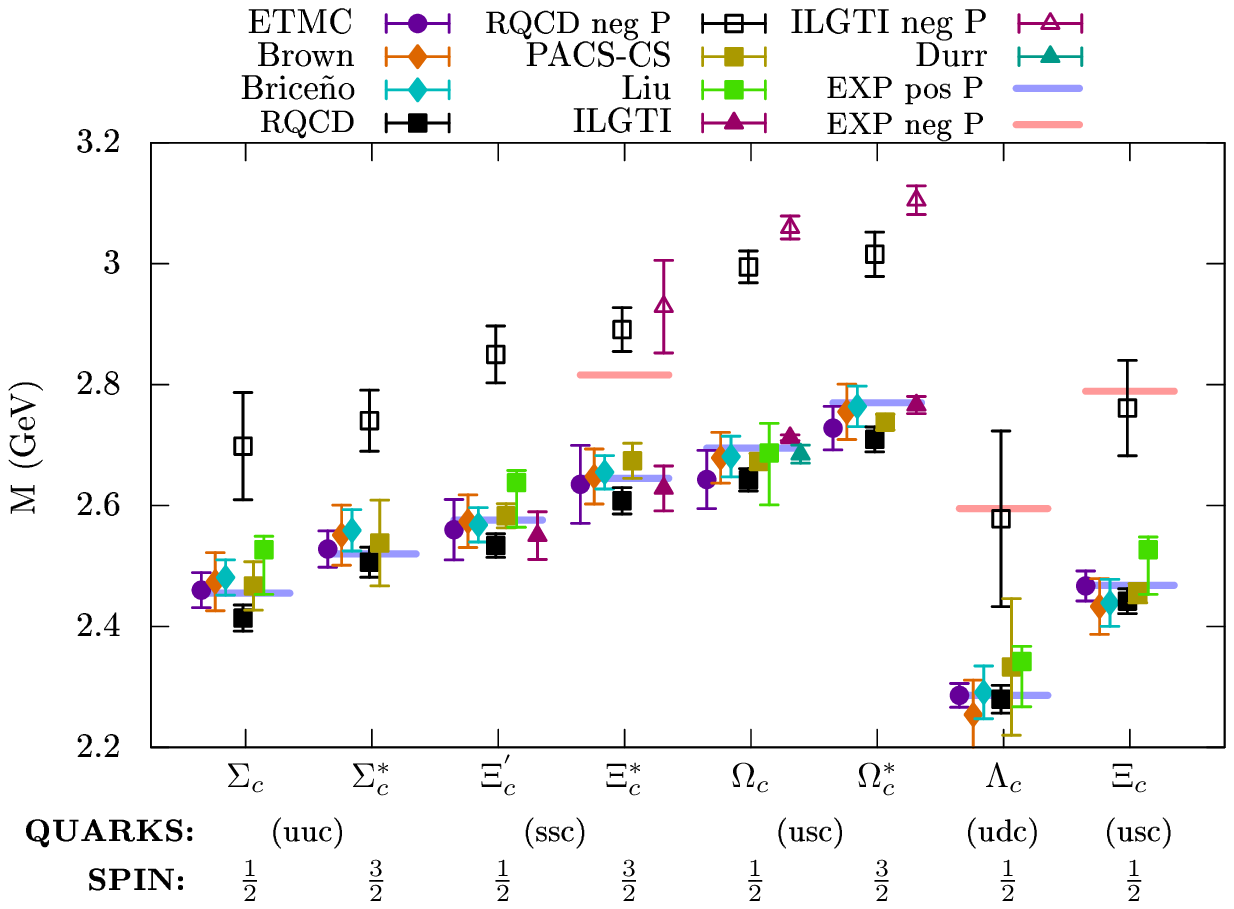}
\caption{Compendium (Ref. \cite{Bali:2015lka}) of low lying singly charmed baryon states from ETMC \cite{Alexandrou:2014sha}, Brown 
{\it et al}., \cite{Brown:2014ena}, Brice\~no {\it et al}., \cite{Briceno:2012wt}, RQCD \cite{Bali:2015lka}, 
PACS-CS \cite{Namekawa:2013vu}, Liu {\it et al}., \cite{Liu:2009jc}, ILGTI \cite{Basak:2013oya} and D\"urr 
{\it et al}., \cite{Durr:2012dw}. The results for a given baryon are ordered along the horizontal direction
such that moving from right to left the estimates possess better knowledge of its systematic uncertainties. Details of 
the systematics addressed in different calculations are briefed in the text.}
\eef{singlyground}

\bef
\small
\parbox{.5\linewidth}{
\centering
  \incfig{0.59}{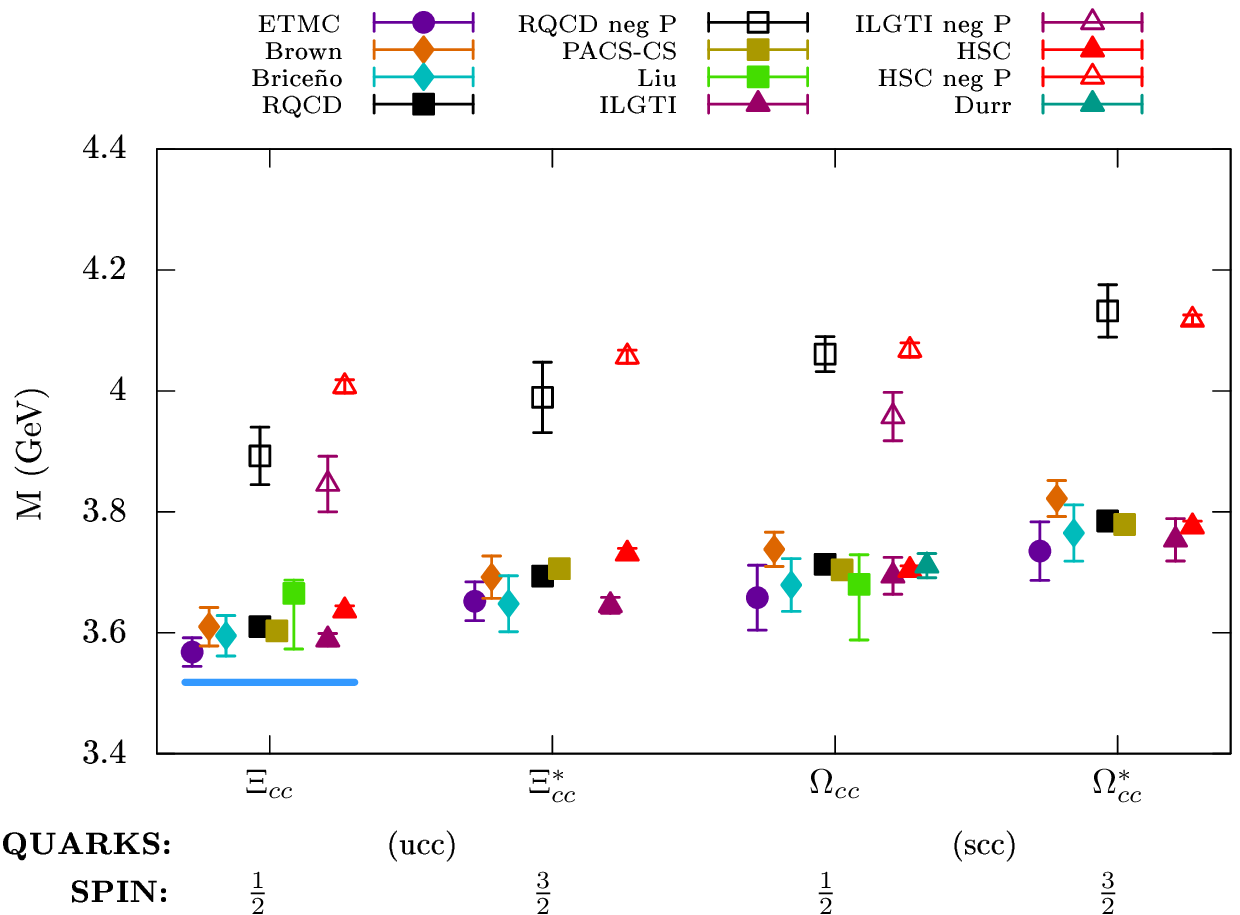}}
\parbox{.5\linewidth}{ 
\centering
  \incfig{0.59}{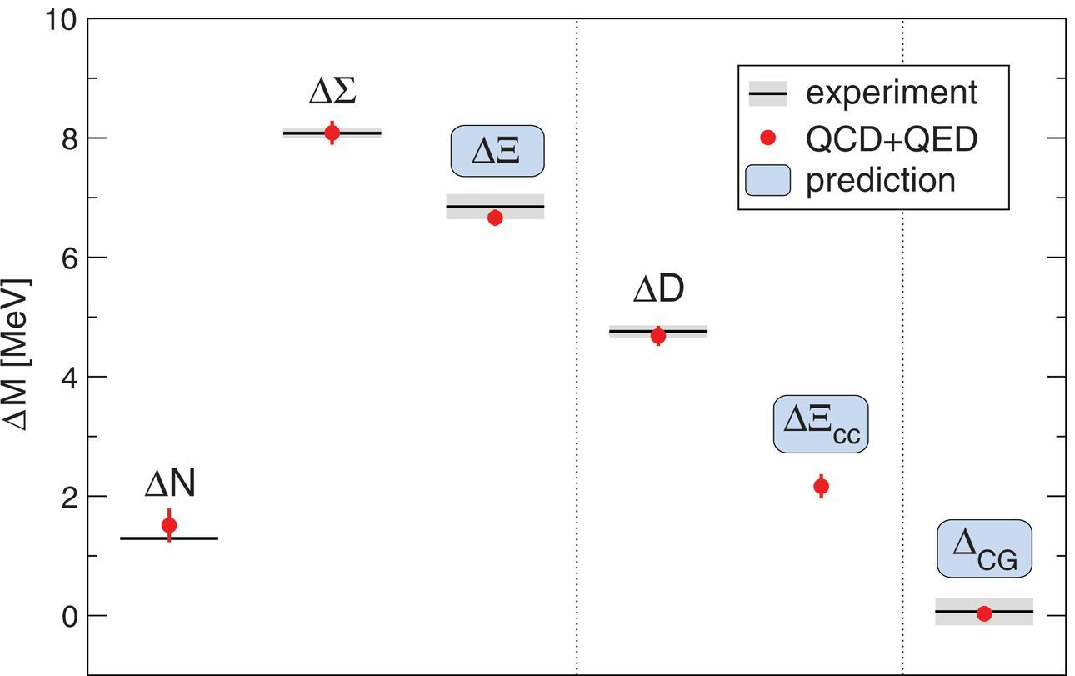}\\}
\caption{Left : Summary (Ref. \cite{Bali:2015lka}) of the low lying doubly charmed baryon states. Details are the 
same as in \fgn{singlyground}. Right : Isospin mass splittings of light and charmed hadrons that are stable under 
strong and electromagnetic interactions by BMW collaboration \cite{Borsanyi:2014jba}. Details in the text. }
\eef{ccq}

In Fig. (\ref{fg:singlyground}), a summary of the recent lattice results for the singly charmed baryon ground 
states has been shown. The results have been ordered along the horizontal direction for each charmed baryon such 
that going from right to left the estimation possess better knowledge of its systematic uncertainties. Of all 
the lattice systematics, chiral and continuum extrapolation are the most important for the charmed baryon ground
states. Starting from left : violet circle ETMC \cite{Alexandrou:2014sha} is a fully dynamical calculation on 
2+1+1 flavor gauge ensembles using twisted mass fermion discretization, with chiral and continuum extrapolated
results; brown diamond (Brown {\it et al}. \cite{Brown:2014ena}) and cyan diamond (Brice\~no {\it et al}., 
\cite{Briceno:2012wt}) are mixed action calculations with chiral and continuum extrapolated estimates. These 
chiral extrapolations follow Heavy Hadron Chiral Perturbation Theory (HH$\chi$PT) allowing not only the 
lattice determination of the spectrum but also the low energy constants from the effective field 
theory. For other numbers in the plot, either one of these two systematics has not been addressed. However, a 
consistency between all the lattice estimates gives confidence and with the increasing computing infrastructure 
\begin{wrapfigure}{l}{85mm}
  \incfig{1.8}{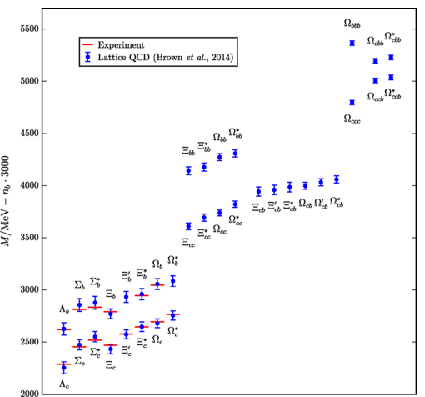}
\caption{Summary of the lattice determination of charm and/or bottom baryons (along with experimental candidates, 
whichever known.) by Brown {\it et al} \cite{Brown:2014ena}. Masses of baryons with $n_b$ bottom quarks are shown 
with an offset of $-n_b ~. ~3 ~GeV$. }
\label{fg:brown}
\end{wrapfigure}
the systematic uncertainties will be pinned down to a level of experimental precision.

In the left of Fig. (\ref{fg:ccq}), a summary of the recent lattice results for the doubly charmed ground states 
are shown. The details of the systematics are the same as that in Fig. (\ref{fg:singlyground}). All 
lattice estimates consistently predict the position of the low lying $\Xi_{cc}$ baryon to be $\sim$80 MeV above the experimentally 
claimed value \cite{Mattson:2002vu}. This raises questions about its discovery. On the right of Fig. (\ref{fg:ccq})
we show the fully controlled 1+1+1+1 flavor ab initio QCD+QED lattice determination (BMW \cite{Borsanyi:2014jba})
of the isospin mass splittings for low lying light and charmed hadrons that are stable under strong and electromagnetic 
interactions. A value somewhat different from the physical isospin splitting for nucleons would have resulted in a 
completely different universe from that of ours. The precise determination of the nucleon isospin splitting with fully 
controlled systematic uncertainties makes this calculation unique. They estimate the isospin splitting for the ground 
state $\Xi_{cc}$ baryons to be 2.16(11)(17) MeV. This again raises questions about the validity of the observations of 
doubly charmed baryons by SELEX with isospin splittings $\sim$17 MeV. More results from these ensembles are much awaited.

With the advent of LHCb, bottom and charmed-bottom baryons also begin receiving significant prospects. A comprehensive lattice 
study of charmed and/or bottom baryons were made in the past on lattices with pure gauge action \cite{Mathur:2002ce}. 
In Fig. (\ref{fg:brown}), we show a summary of all the charmed and/or bottom baryons as estimated from 
a mixed action lattice calculation with controlled systematics by Brown {\it et al} \cite{Brown:2014ena}. 
In the left of Fig. (\ref{fg:ccc}), we show the lattice determinations for the ground state triply charmed baryon 
with an offset of 3/2 times the mass of $J/\psi$ meson from respective calculations.

\vspace{-0.3cm}
\section{Excited state spectroscopy}
\vspace{-0.3cm}
\bef[h]
\small
\parbox{.5\linewidth}{
\centering
  \includegraphics[scale=0.3]{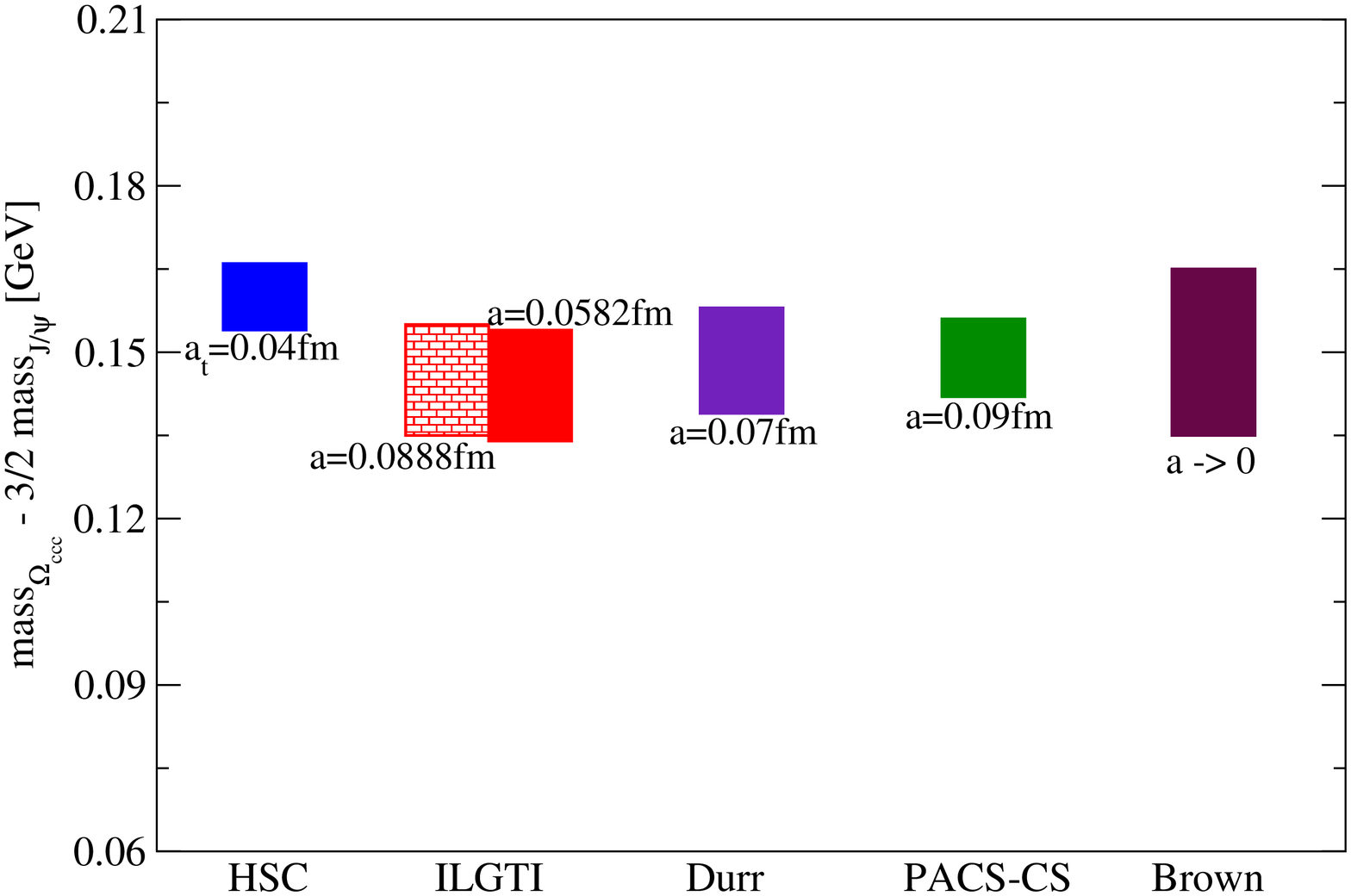}}
\parbox{.5\linewidth}{ 
\centering
  \includegraphics[scale=0.3]{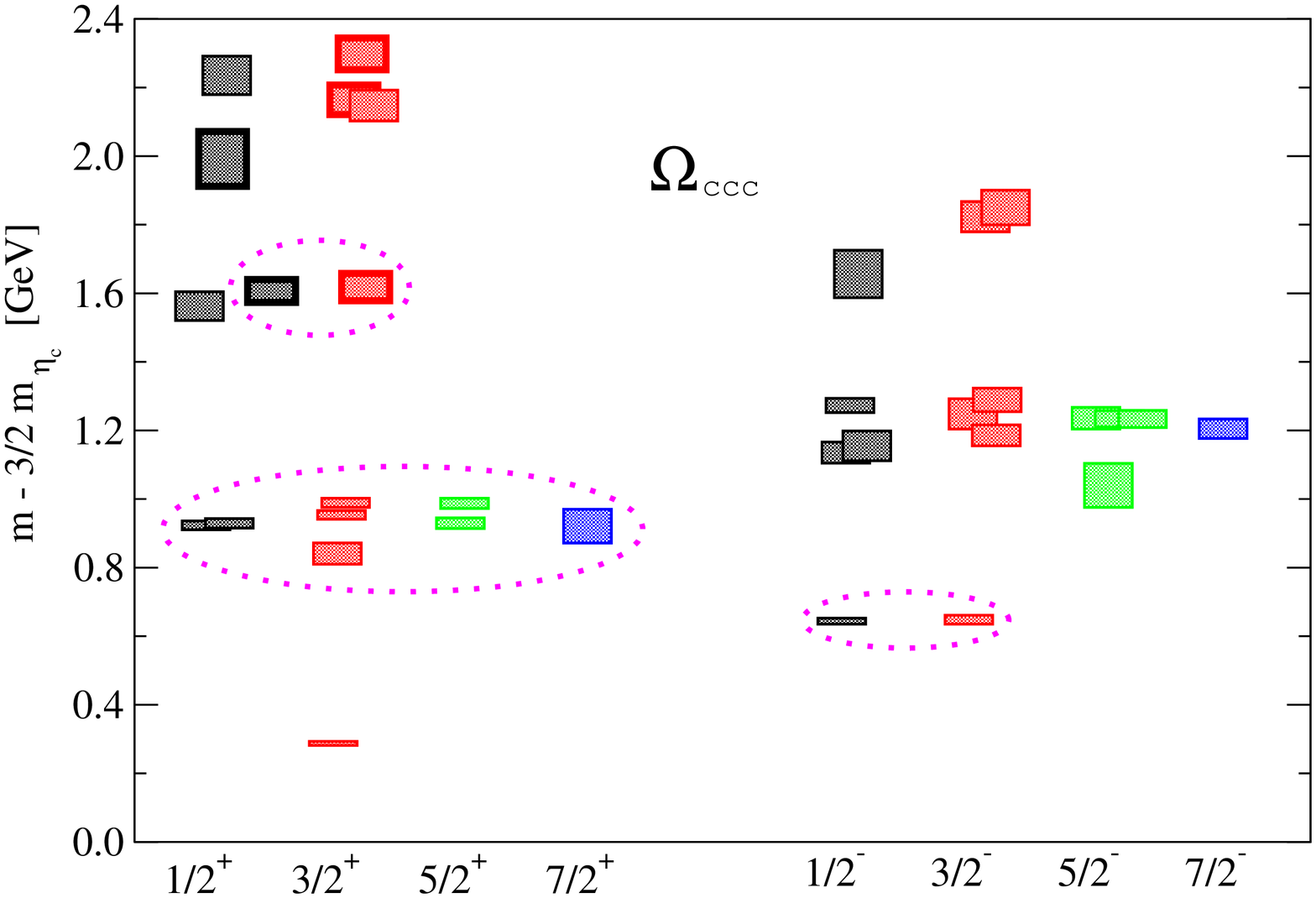}}
\caption{ Left : Summary of lattice calculations of ground state triply charmed baryons from HSC \cite{Padmanath:2013zfa}, 
ILGTI \cite{Basak:2013oya}, D\"urr \cite{Durr:2012dw}, PACS-CS \cite{Namekawa:2013vu} and Brown {\it et al} 
\cite{Brown:2014ena}. Right : Spin identified spectra of triply-charmed baryons with respect to $\frac32 m_{\eta_c}$ \cite{Padmanath:2013zfa}. 
The boxes with thick borders correspond to the states with strong overlap with hybrid operators. The states inside 
pink ellipses are those with relatively large overlap to non-relativistic operators.}
\eef{ccc}

The excited state information lies in the sub-leading exponential of the spectral decomposition (\eqn{2pt}) of 
the two point correlation functions. Extraction of these observables are extremely unstable by 
conventional spectroscopy techniques. This limits the lattice baryon investigations to the ground states 
with spin up to 3/2 \cite{Bali:2015lka,Alexandrou:2014sha,Brown:2014ena,Briceno:2012wt,Namekawa:2013vu,Liu:2009jc,Basak:2013oya,Durr:2012dw}, 
until very recently where a comprehensive spectra of charmed 
baryons have been presented \cite{Padmanath:2013zfa,Padmanath:2015jea}. The calculation proceeds through computation of matrix of 
two point correlation functions with a basis of carefully constructed charmed baryon operators using a 
derivative-based operator construction formulation \cite{Edwards:2011jj}. The operators are constructed 
such that they are expected to probe the radial as well as orbital excitations in the respective systems. 
Quark fields are also smeared to suppress the high frequency modes and to improve the overlap onto the desired 
low lying states. Extraction of the excited state information from the matrix of correlation 
functions thus computed is performed by solving the generalized eigenvalue problem, which expresses the physical 
states as a linear combination of the used set of interpolators. Thus using a very large basis of 
interpolating operators with widely different spatial structure, one can extract the ground and 
the excited states very reliably. Such calculations have been established in studying the light baryon spectrum \cite{Edwards:2011jj,Edwards:2012fx}.

\bef[h]
\small
\parbox{.5\linewidth}{
\centering
  \includegraphics[scale=0.3]{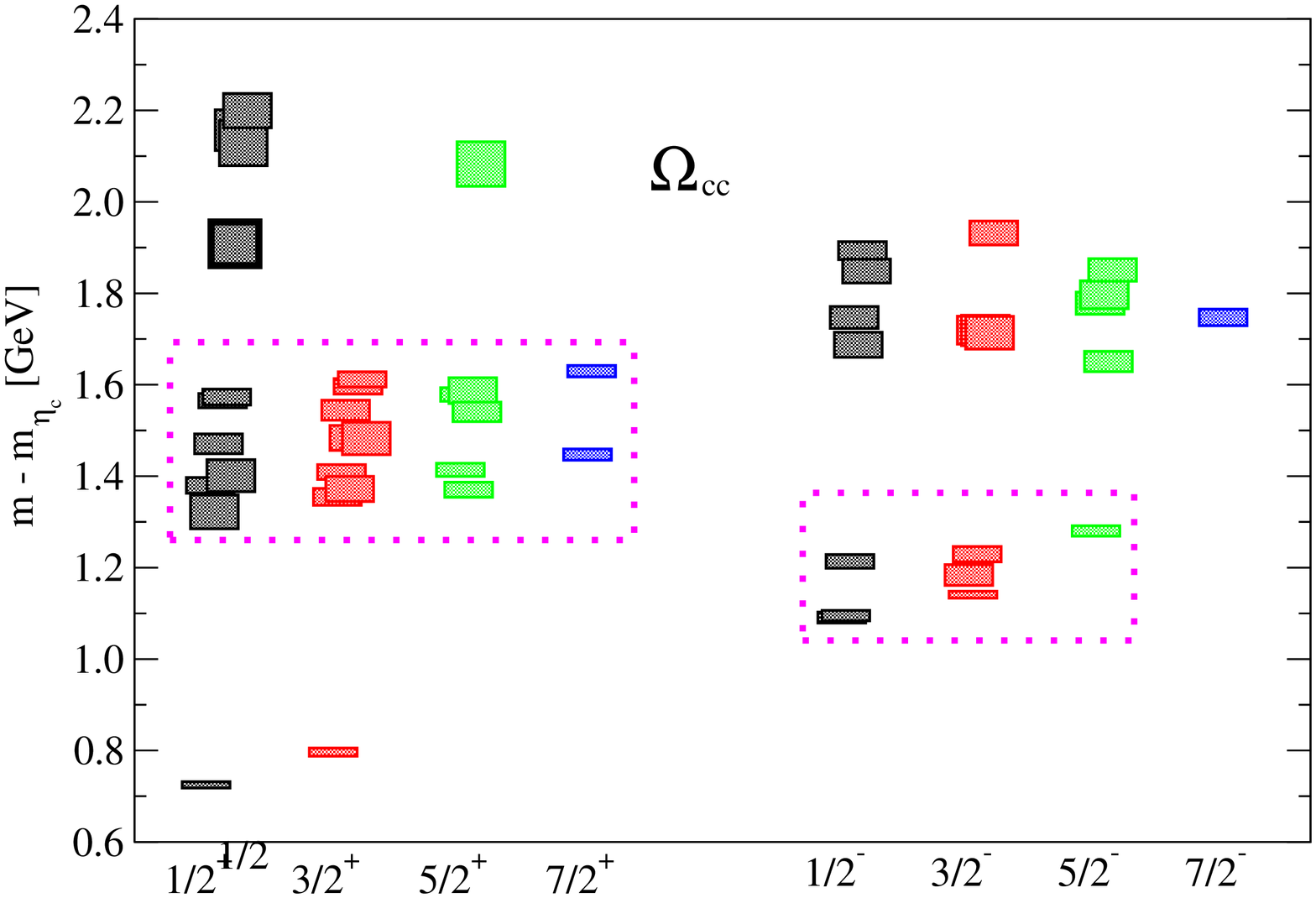}\\
(a)}
\parbox{.5\linewidth}{ 
\centering
  \includegraphics[scale=0.3]{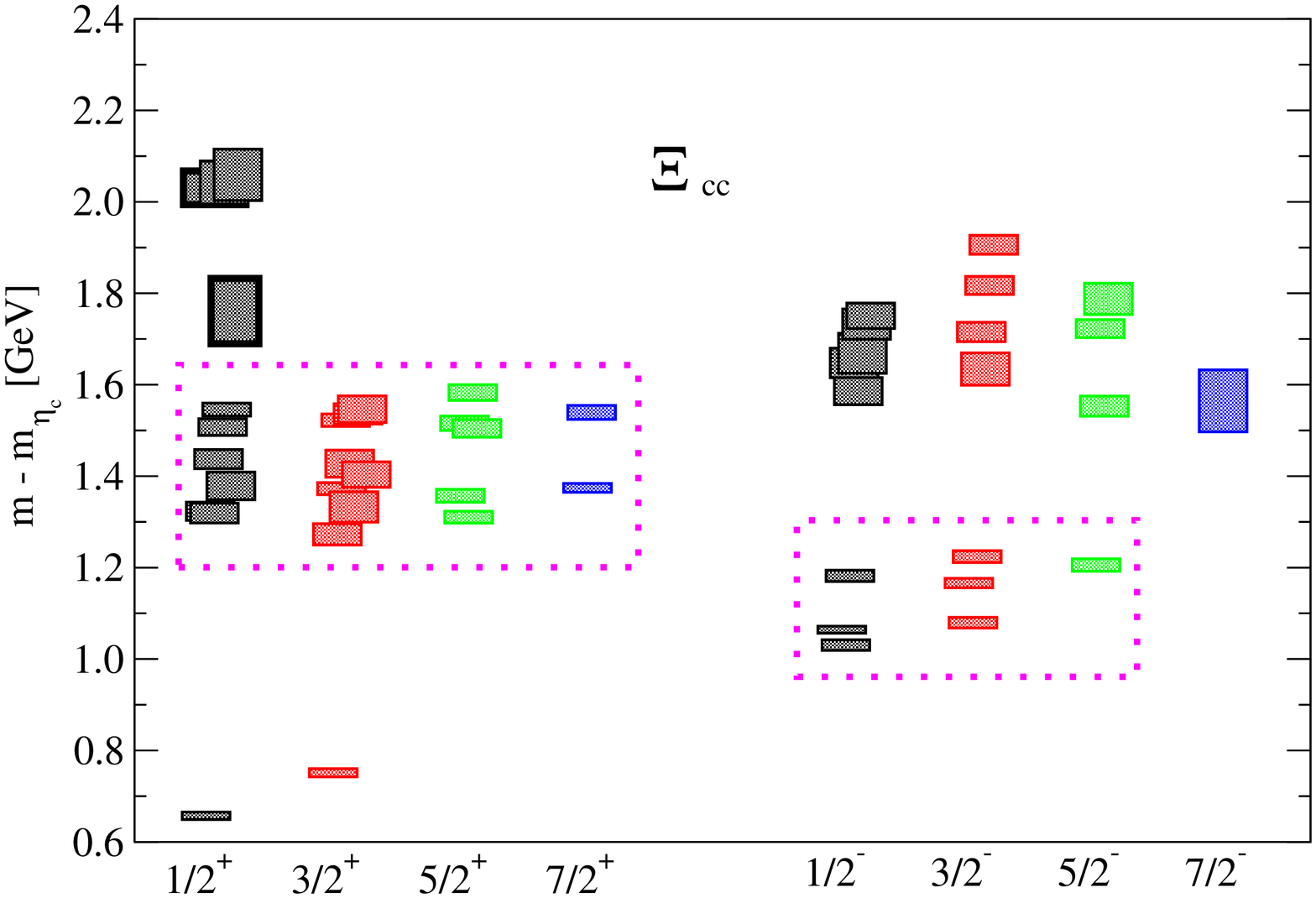}\\
(b)}
\caption{ Spin identified spectra of (a) $\Omega_{cc}$ and (b) $\Xi_{cc}$ baryon for both parities 
and with spins up to $\frac72$ {\it w.r.t.} $m_{\eta_c}$ \cite{Padmanath:2015jea}. The keys are same as in right of \fgn{ccc}.}
\eef{ccq_spectrum}

In the remaining part, we focus on the recent lattice calculation of excited charmed baryon spectra by 
Hadron Spectrum Collaboration (HSC) \cite{Padmanath:2013zfa,Padmanath:2015jea}. With as many as 90 different interpolators 
for each charmed baryon, the computation of the correlation functions itself can be computationally as
challenging as gauge field generation. It is to be mentioned that these calculations of excited charmed 
baryons are exploratory and currently lack a good control over the systematic uncertainties. Nevertheless, 
these calculations serve a pioneering step towards understanding baryon resonances and baryon-meson scattering. 
For these studies, we utilized the dynamical anisotropic 
gauge-field configurations generated by the Hadron Spectrum Collaboration (HSC) 
to extract highly excited hadron spectra. The temporal lattice spacing is $a_t^{-1}=5.67$GeV and lattice 
spatial extension is $L=1.9$ fm, which presumably is sufficiently large for charmed baryon spectroscopy. 
The pion mass on these lattices is 391 MeV. Further details of these lattices can be found in Refs. 
\cite{Edwards:2008ja, Lin:2008pr}.

In the right of Fig. (\ref{fg:ccc}) we show the spin identified spectra of the triply charmed baryons,
in terms of splitting from $3/2$ times $m_{\eta_c}$ to account for the difference in the charm quark 
content \cite{Padmanath:2013zfa}. Energy splittings are in general preferred, as it reduces various systematic 
uncertainties. Boxes with thicker borders correspond to those states with dominant overlap onto the 
operators that are proportional to the field strength tensor, which might consequently be hybrid 
states. We classify the states within the magenta ellipses to be dominantly non-relativistic in nature
due to their relatively large overlap with non-relativistic operators. One important observation is that 
even though we use a full relativistic plus non-relativistic set of operators, the pattern of the low 
lying bands exactly agree with expectations from models with an $SU(6)\otimes O(3)$ symmetry. Similar 
calculation of triply bottom baryons has been reported in Ref. \cite{Meinel:2012qz}.

\bef[h]
\vspace*{-0.5cm}
\small
\parbox{.5\linewidth}{
\centering
  \includegraphics[scale=0.3]{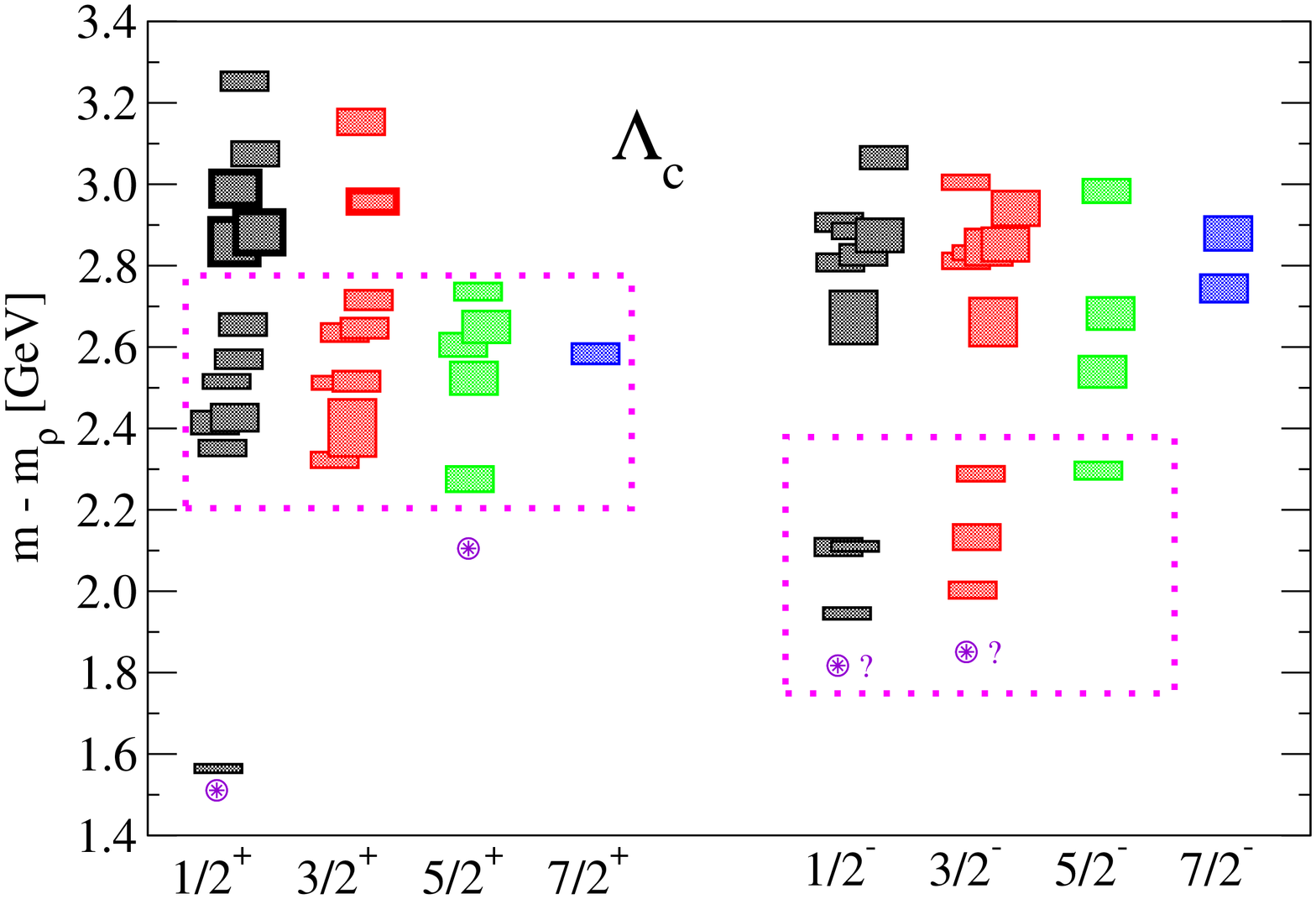}\\
(a)}
\parbox{.5\linewidth}{ 
\centering
  \includegraphics[scale=0.3]{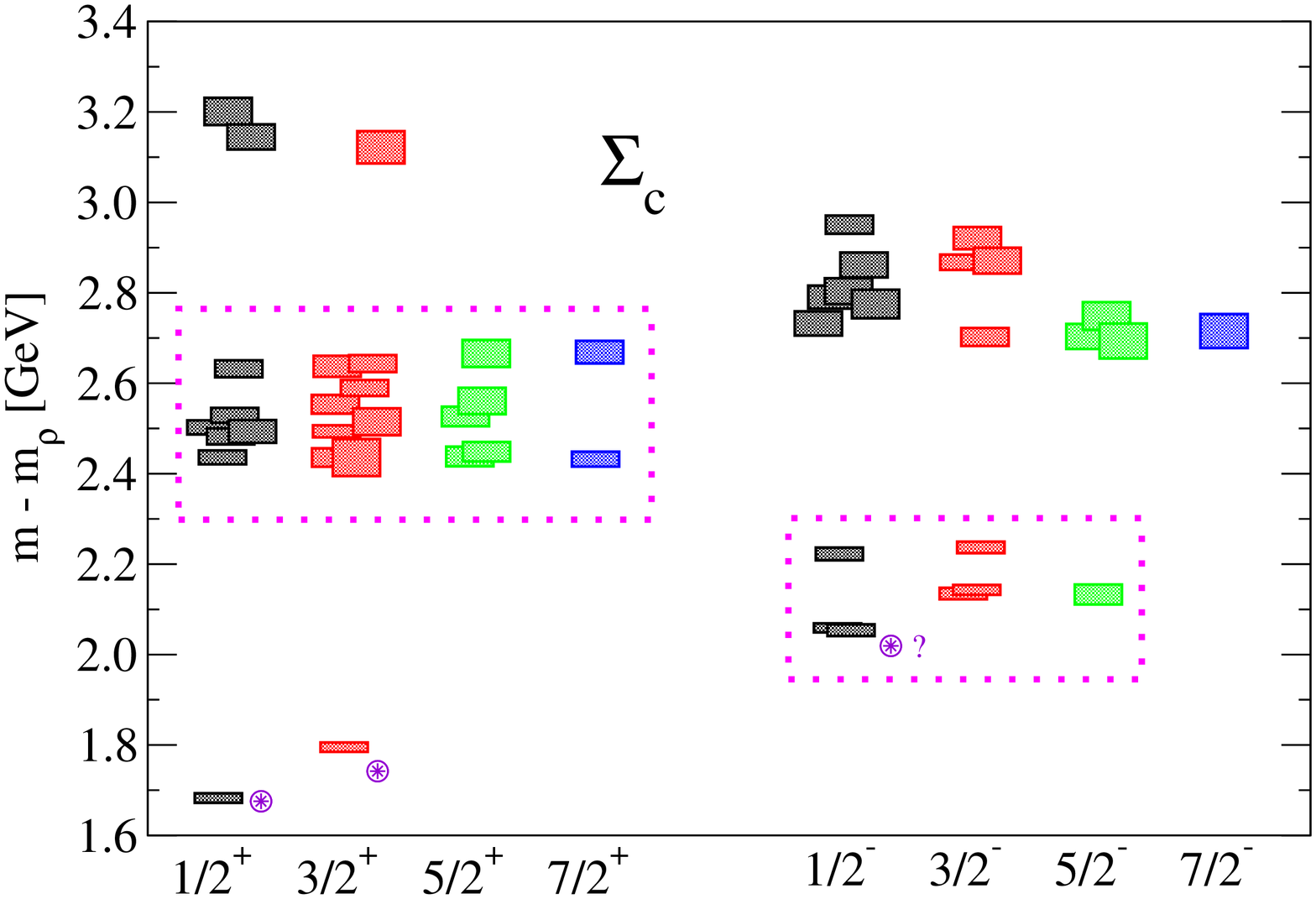}\\
(b)}
\parbox{.5\linewidth}{
\centering
  \includegraphics[scale=0.3]{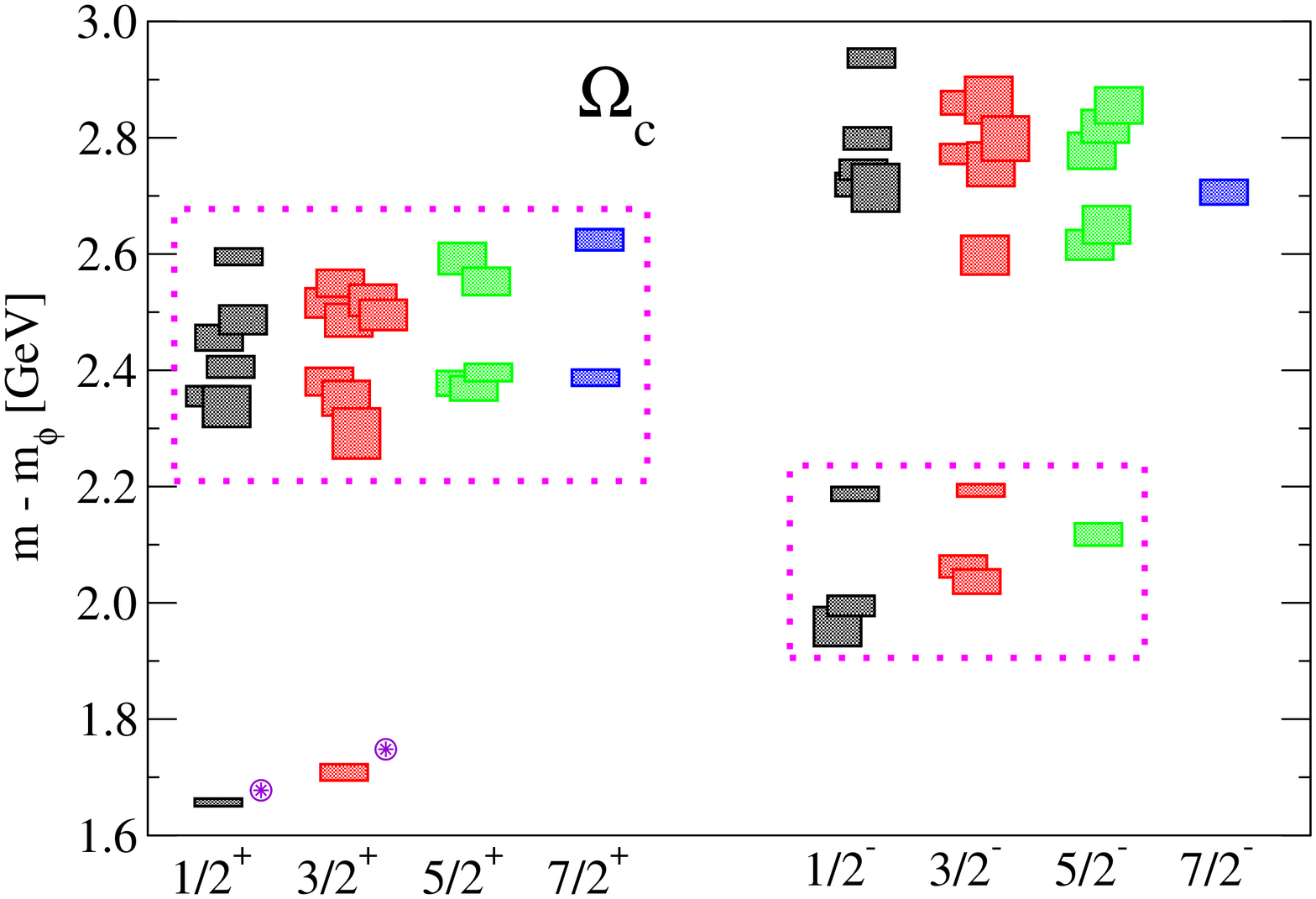}\\
(c)}
\parbox{.5\linewidth}{ 
\centering
  \includegraphics[scale=0.3]{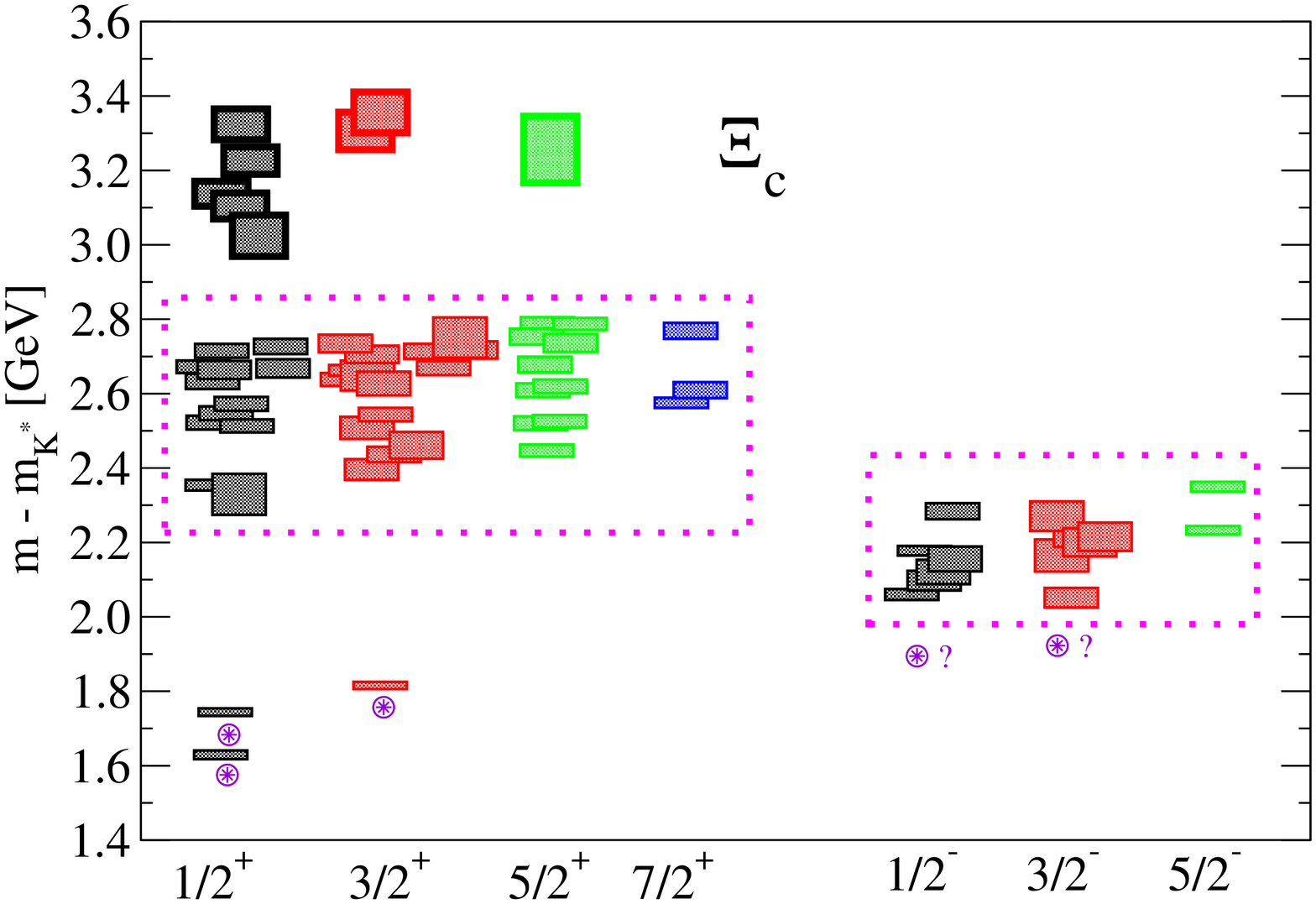}\\
(d)}
\caption{The spin identified spectra of (a) $\Lambda_{c}$, (b) $\Sigma_{c}$, 
(c) $\Omega_{c}$ and (d) $\Xi_{c}$ baryons for both parities in terms of the splitting from 
the respective vector mesons. The keys are same as in the right of \fgn{ccc}.}
\eef{LS_spectrum}

\fgn{ccq_spectrum} shows the spin identified excited doubly charmed baryon spectra \cite{Padmanath:2015jea} as the difference 
from $m_{\eta_c}$ to account for the difference in the charm quark content. The boxes with thick borders again represent 
candidates for doubly charmed baryons with strong gluonic excitations, while the states within the magenta rectangles 
are classified as non-relativistic in nature based on their dominant spectral overlaps. Identification and classification 
of higher lying states in the third excitation band and higher are not well-defined, since the set of interpolators we use 
in our studies are limited to a maximum two derivatives. Once again the number and pattern of the low lying bands 
of the excited state spectra show consistency with prediction based on a model with SU(6)$\otimes$O(3) symmetry.
The excited spectra of singly charmed baryons ($\Lambda_c$, $\Sigma_c$, $\Xi_c$ and $\Omega_c$) are as shown in \fgn{LS_spectrum}.
The spectra for singly charmed baryons are shown as the splitting from respective light vector meson mass, so that all 
the plots in \fgn{LS_spectrum} have same remnant valence charm quark content. A few of the low lying experimental candidates are 
also shown as violet circled stars in the respective plots for comparison. Those experimental candidates with uncertain 
quantum numbers are identified with a question mark `?' within the plots. The singly charmed baryon spectra also show
good agreement with expectation as per a model with SU(6)$\otimes$O(3) symmetry. 

\vspace{-0.3cm}
\section{Conclusions}

The wealth of experimental and theoretical prospects in heavy baryon spectroscopy calls for urgent scientific efforts 
from both the fronts. In addition, in a recent lattice calculation at finite temperature, signatures for existence of 
many additional charmed baryon excitations, {\it w.r.t} what is known from experiments, have been observed.
First principles 
calculation, such as lattice QCD, can play a significant role in predicting many charmed baryon excitations yet to be 
observed in experiments, in revealing the role of diquark correlations in baryons, the freezing degrees of freedom, etc. 

Lattice calculations of the low lying charmed baryon states have entered an era of precision spectroscopy with multiple 
collaborations making full QCD simulations of charmed baryon ground states with full control over the systematic 
uncertainties. Compendium of lattice determinations of singly, doubly and triply charmed baryons are as shown in 
Figs. \ref{fg:singlyground}, \ref{fg:ccq} and \ref{fg:ccc}. \fgn{brown} shows the summary of charmed, bottom and charmed-bottom
baryons from a mixed action calculation reported in Ref. \cite{Brown:2014ena} with controlled systematics. The first 
fully controlled QCD+QED 1+1+1+1 flavor lattice determination of isospin splitting of the doubly charmed baryon by BMW
collaboration \cite{Borsanyi:2014jba} deserves special mention. They estimate this value to be 2.16(11)(17) MeV, which 
raises questions against the only doubly charmed candidate observed.

The first exploratory calculation of excited charmed baryon spectra has been made in Refs. \cite{Padmanath:2013zfa,Padmanath:2015jea} and a controlled systematic study 
of excited charmed baryons is on the run. The calculations proceeds through computation of matrices of two point 
correlation functions of large set of interpolators and expressing the physical states as a spectral combination of 
these interpolators. Figs. \ref{fg:ccc}, \ref{fg:ccq_spectrum} and \ref{fg:LS_spectrum} show the excited spectra of 
charmed baryons from these calculations. The ground states from these calculations are also compared to find agreement 
with other lattice determinations in Figs. \ref{fg:ccq} and \ref{fg:ccc}. The excited spectra that we obtain 
have excitations with well-defined spins up to 7/2 for both the parities and the low lying levels resemble the 
expectations from a model with $SU(6)\otimes O(3)$ symmetry. However, with as many as 90 interpolators for each 
charmed baryons, the computation of the matrices of correlation functions are computationally as demanding as gauge 
field generation. It is to be mentioned that these calculations are at an exploratory level and the systematic 
uncertainties like chiral and continuum extrapolation and infinite volume limits have not been addressed here. 
However, it serves as a pioneering work towards understanding baryon resonances. Efforts are being made to include 
baryon-meson kind of operators to account for the possible scattering and decay of the excitations. The absence of 
such operators may affect some of the above conclusions, however to a lesser extent than their influence in the 
light hadron spectra. 

A number of lattice calculations addressing the electromagnetic form factors and radiative transitions of charmed baryons,
semi-leptonic form factors and exclusive decay rates of $\Lambda_b$ baryons exist in literature. These 
subjects have not been covered in this review due to space constraint and interested readers can find the details in Refs. \cite{Can:2013tna,Meinel:2014wua}.

\vspace{-0.3cm}
\section{Acknowledgements}
We would like to thank our collaborators Robert Edwards and Mike Peardon. We also thank Paula Perez-Rubio and Sayantan 
Sharma for providing us material for the talk. We would also like to acknowledge Christian Lang and Stefan Meinel 
for discussions. Computations related to our work were done at gaggle and brood clusters of DTP,TIFR, computational 
facilities of ILGTI, Trinity college as well as at Jefferson Lab. NM acknowledges support 
from the Department of Theoretical Physics, TIFR, and MP acknowledges support from the Austrian Science Fund (FWF):[I1313-N27].


\end{document}